\documentclass[10pt]{article}
\usepackage[cp1251]{inputenc}
\usepackage[english]{babel}
\usepackage{graphicx}
\title{{\bf \Large Cosmological Sigma Model with Non-Minimal\\ Coupling to the Target Space }\\
\vspace{3mm}{\normalsize ~~{\bf V.\,K. Shchigolev}\thanks{E-mail: vkshch@yahoo.com}\\\vspace{3mm}
{\small {\it Department of Theoretical Physics, Ulyanovsk State University, Ulyanovsk 432000, Russia}}\\
\vspace{3mm}
\small \begin{quote}{\bf Abstract} --  A homogeneous
and isotropic Universe in the framework of nonlinear sigma model with non-minimal coupling to the target space is considered. A two-component model of such a sort is preliminary investigated.    Some solutions for this model are given. Perspectives and directions of development of such a sort  of models are designated.\\
\vspace{2,5mm}
PACS numbers: 98.80.-k, 95.36.+x, 02.40,11.10.Ef\\
Key words: Sigma model, chiral fields, cosmology, non-minimal coupling, target space.\\
\end{quote}}}

\date{}

\begin{document}

\maketitle \vspace{-2cm}

\section {\large Introduction}

\qquad As known, all cosmological observations of the last time
(see, e.g. \cite{C1}-\cite{C4}) strongly indicate that the observable Universe is undergoing a phase of accelerated expansion. There are at least two ways to solve the problem of late-time acceleration, namely the modification of the gravitational theory and the introduction of some exotic source of gravity, which is commonly called Dark Energy (DE).
As the models of dark energy  are customarily used not only canonical fields with the modified potential of self-interaction or the combination of fields with some physically or phenomenologically founded   types of interaction between the components \cite{C5}-\cite{C8}). The whole direction in the problem of dark energy is devoted to the study of fields with non-canonical kinetic term \cite{C9}-\cite{C16}, among which we could mention the study of the kinetic terms of the type $K=(1/2)f(\phi)g^{ik}\nabla_i \phi \nabla_k \phi$. Kinetic terms of  a similar type arise naturally in the nonlinear sigma model (NSM),  where the functions $f_A(\phi)$ with $A=1,2,..,N$ are defined by means of metric coefficients of  the chiral target space

A chiral cosmological model as NSM with a potential
has been already used as a model of the very early
Universe and inflation. As shown in \cite{C17}, this type of cosmological model can be successfully used in solving the problem of the late-time acceleration with a chiral Dark Sector as well.  This kind of fields  has been extensively
applied to the DE problem in the modern cosmology
\cite{C18}-\cite{C23}.
Chiral NSM were introduced as a theory of strong interactions by Schwinger \cite{C24} and
Skyrme \cite{C25}. To our knowledge, consideration of NSM as the source of the
gravitational field was proposed by G. Ivanov \cite{C26}. Later on, the cosmological applications of NSM is developed by S.V. Chervon (see e.g. \cite{C27}, \cite{C28} and references therein) and some other authors.
The idea of non-minimal coupling and interaction of the matter components of  the universe has become very attractive in the recent years due to the numerous attempts to solve the mystery of the Dark Sector. An interesting feature of chiral cosmological models is that the interaction between the components in their framework can be described geometrically by means of the metric coefficients of the target space.

As a rule, all modifications of the kinetic terms in cosmological sigma models are entirely determined  by the internal geometry of the chiral (target) space. Any modification of of kinetic terms, derived from some additional requirements on the target-space  geometry,  could be considered as a phenomenological approach. Thus, these modifications already very few differ from the  models with the multiplets of scalar fields which  have repeatedly been  investigated (see, for example, \cite{C29}-\cite{C31}).

However, there is a new modification of the kinetic terms from the same internal geometry of the target space.  It is associated with the fact that the kinetic terms of the chiral fields $K=(1/2)h_{AB}(\phi^C)g^{ik}\nabla_i \phi^{A}\nabla_k \phi^{B}$ may contain not only a chiral metric $h_{AB}(\phi^C)$, but also some tensors of the second rank built on the Ricci tensor, a combination of convolutions with the curvature tensor of the target space,  and so on. This type of self-interaction of the chiral fields can be called a non-minimal coupling to the target space in the total Lagrangian.  In this case, modification of the kinetic term is naturally based on the same internal geometry, and does not require any artificial or phenomenological assumptions. It is interesting to investigate how such a modification is able to enrich the theory,  and what are its limits in converting kinetic terms. We are going to carry out a preliminary study of this problem.

\section {\large Model and the Field Equations}

\quad  Consider a non-linear sigma model with non-minimal coupling of the chiral fields to the target space $\{{\cal N},h_{AB}(\phi)\}$,
assuming that the action is as follows:
\begin{equation}
\label{1} S_{n-m}^{\,\sigma}=\int_{\cal M}
\left(\frac{R}{2\kappa^2}+\frac{1}{2}{\cal H}_{AB}\phi_i^A \phi_k^B
g ^{ik}-V(\phi^C)\right)\sqrt{-g}~ d^4 x ,
\end{equation}
where $\kappa^2 = 8 \pi G = M_P ^{-2}$ is the gravitational
constant, $M_P$ is the Planck mass, $\{{\cal M},g_{ik}(x)\}$ is a
space-time, $\{{\cal N},h_{AB}(\phi)\}$ is a chiral space of the
fields  $\phi^A = (\phi ^1,...,\phi^n)$,  $\phi_{,i}^A\equiv \partial_i\phi^A$,  $g=\det (g_{ik})$,  and $V(\phi^C)$ is a potential of
the chiral fields.
The simplest non-minimal coupling to the target space implies that
\begin{equation}
\label {2} {\cal H}_{AB} =\alpha\, h_{AB}+\beta \,{\cal R}\, h_{AB} + \gamma\,
{\cal R}_{AB},
\end{equation}
where $\alpha, \beta$ and $\gamma$ are dimensional coupling constants,
${\cal R}_{AB}$ is the Ricci tensor built on the metric of target space $h_{AB}$:
\begin{equation}\label{3}
{\cal R}_{AB}=\frac{\partial \Gamma^{C}_{AB}}{\partial \phi^C}-\frac{\partial \Gamma^{C}_{AC}}{\partial \phi^B}+\Gamma^{C}_{AB}\Gamma^{D}_{CD}-\Gamma^{D}_{AC}\Gamma^{C}_{BD},
\end{equation}
with the help of the Christoffel  symbols given by
\begin{equation}\label{4}
\Gamma^{A}_{BC}=\frac{1}{2}h^{AD}\Big(\frac{\partial h_{DC}}{\partial \phi^{B}}+\frac{\partial h_{BD}}{\partial \phi^{C}}-\frac{\partial h_{BC}}{\partial \phi^{D}}\Big),
\end{equation}
and ${\cal R} = h^{AB}{\cal R}_{AB}$ is the curvature scalar of the target space.

The set of dynamical equations corresponding to the action (\ref{1}) is as follows:
\begin{equation}\label{5}
R_{ik}-\frac{1}{2}g_{ik}R = \kappa^2 T_{ik},
\end{equation}
where the energy-momentum tensor (EMT) is given by
\begin{equation}\label {6}
T_{ik}={\cal H}_{AB}\phi_{,i}^A \phi_{,k}^B -g_{ik}\Big[\frac{1}{2}{\cal H}_{AB}\phi_{,m}^A \phi_{,n}^B g^{mn} - V(\phi^C)\Big].
\end{equation}
Variation of the action (\ref{1}) with respect to the fields $\phi^A$ yields
the following set of equations
\begin{equation}
\frac{1}{\sqrt{-g}}\frac{\partial}{\partial x^i}\left(
\sqrt{-g}\,{\cal H}_{AB}g^{ik}\phi^B_{,k}\right)-\frac{1}{2}\,\frac{\partial{\cal H}_{BC}
}{\partial \phi^A}\phi^B_{,i}\phi^C_{,k}g^{ik}
+\frac{\partial V}{\partial\phi^A} = 0. \label{7}
\end{equation}
This equation can be rewritten in a different form by introducing the coefficients of the connection on the redefined target-space metric  (\ref{2}),
\begin{equation}\label{8}
{\cal T}^{A}_{BC}=\frac{1}{2}{\cal H}^{AD}\Big(\frac{\partial {\cal H}_{DC}}{\partial \phi^{B}}+\frac{\partial {\cal H}_{BD}}{\partial \phi^{C}}-\frac{\partial {\cal H}_{BC}}{\partial \phi^{D}}\Big),
\end{equation}
as follows:
\begin{equation}
\frac{1}{\sqrt{-g}}\frac{\partial}{\partial x^i}\left(
\sqrt{-g}\,g^{ik}\phi^A_{,k}\right)+{\cal T}^{A}_{BC}\,\phi^B_{,i}\phi^C_{,k}g^{ik}
+\frac{\partial V}{\partial\phi^B}{\cal H}^{AB} = 0. \label{9}
\end{equation}
It is easy to see that even in the cases of a linear function of the potential $V(\phi^C)$ on  the chiral fields or even its constancy, the equation (\ref{9}) is nonlinear in the second term. This leads to the nonlinear chiral model and the nonlinear sigma model in cosmology. This equation contains the second term which is induced by the intrinsic geometry of the target space.
In the standard theory of self-interacting
scalar field, a similar non-linear term could appear mostly due to the phenomenological approach.

For the sake of simplicity, we consider a flat Friedman-Robertson- Walker (FRW) model with the
space-time interval
\begin{equation}\label{10}
ds^2 = dt^2 - a^2(t)\delta_{ik}\,dx^i \,dx^k,
\end{equation}
where $a(t)$ is a scale factor of the Universe. We consider the homogeneous chiral fields $\phi^A (t)$. Due to the latter and metric (\ref{10}), the Einstein equation (\ref{5}) with the EMT (\ref{6}), and the equation for the chiral fields (\ref{9}) take the following form:
\begin{equation}\label{11}
3H^2= M_P ^{-2}\Big[\frac{1}{2}{\cal H}_{AB}\dot\phi^A \dot\phi^B + V(\phi^C)\Big],
\end{equation}
\begin{equation}\label{12}
2\dot H=- M_P ^{-2}{\cal H}_{AB}\dot\phi^A \dot\phi^B,
\end{equation}
\begin{equation}\label{13}
\ddot\phi^A+3H\dot\phi^A+{\cal T}^{A}_{BC}\,\dot\phi^B\dot\phi^C
+\frac{\partial V}{\partial\phi^B}{\cal H}^{AB} = 0.
\end{equation}
This set of equations governs the dynamics of gravity and the chiral fields in the framework of the developed model.

\section{\large A Two-Component Sigma Model}

Following Ref. \cite{C13}, let us consider a two-component sigma model of chiral fields $\phi^1= \phi,\,\,\phi^2=\chi$ with the metric of target space of the form
\begin{equation}\label{14}
d s_{\sigma}^2 =\epsilon d \phi ^2 +  \Sigma^2(\phi) d \chi^2,
\end{equation}
that is, assuming that the non-zero components of the chiral metric $h_{11}=\epsilon,\,h_{22}= \Sigma^2(\phi)$. Here and in what follows, $\epsilon = \pm 1$ used for the canonical and phantom fields $\phi$, respectively. From this we can easily find the following non-zero components ${\cal R}_{AB}$ and ${\cal R}$ in (\ref{2}):
$$
{\cal R}_{11}=\frac{1}{\epsilon\Sigma^2}{\cal R}_{22}=\frac{1}{2 \epsilon}{\cal R}=-\frac{\Sigma''}{\Sigma}.
$$
Here and henceforth, the primes denote derivatives with respect to $\phi$. Thus, the non-minimal metric ${\cal H}_{AB}$ is obtained as
\begin{equation}\label{15}
\epsilon {\cal H}_{22}= \Sigma^2\,{\cal H}_{11}= \epsilon\, Q(\Sigma)\,\Sigma^2\,,\,\,\mbox{where}\,\,\,\, Q(\Sigma)=\alpha-\epsilon \delta\frac{\Sigma''}{\Sigma} ,
\end{equation}
with $\delta=2\beta+ \gamma$. It is seen that non-minimal metric ${\cal H}_{AB} = Q(\Sigma)\,h_{AB}$, that is conformal to the original metric of the target  space  with the factor $ Q (\Sigma) $ if $Q(\Sigma)>0$.
It is easy to find from the definition (\ref{15}), that for a special choice of the coupling coefficients, $\alpha = 1,\,\,   \delta=0\Rightarrow\gamma=-2\beta$,  the non-minimal coupling is absent, i.e. ${\cal H}_{AB}\equiv h_{AB}$. Of course, this is only a consequence of the vanishing of the Einstein tensor ${\cal G}_{AB}={\cal R}_{AB}-(1/2){\cal R}$ or any two-dimensional space. However, this is not the only possibility of degeneration of non-minimal coupling to the target space up to the minimal one. From Eq. (\ref{15}) it follows that  we get the same result, if the target function $\Sigma(\phi)$ satisfies the equation $Q(\Sigma)=1$, that is
\begin{equation}\label{16}
\Sigma\,''+\epsilon\left(\frac{1-\alpha}{\delta}\right) \Sigma=0,
\end{equation}
subject to $\delta \ne 0$. Thus, the non-minimal coupling to the target space degenerates into the usual minimal coupling either when $\delta = 0$ for any function $\Sigma(\phi)$, or, when  $\delta \ne 0$, in the cases following from Eq. (\ref{16}):
$$
\Sigma(\phi) = \left\{
\begin{array}
[c]{rcl}
\sin\left(\displaystyle \frac{\phi}{\phi_0}\right) ~~~\mbox{for}~~\displaystyle \epsilon\frac{1-\alpha}{\delta}>0\,,\\
\phi~~~~~~~~~~~~~~~\mbox{for}~~~\alpha =1\, ,~~~~~~~ &  &  \\
\sinh\left(\displaystyle \frac{\phi}{\phi_0}\right)~~~\mbox{for}~~\displaystyle \epsilon\frac{1-\alpha}{\delta}<0\,,
\end{array}
\right.
$$
where $\phi_0=\sqrt{|\delta/(1-\alpha)|}$, and the particular values of the constants of integration are taken without losing the generality of the result due to the permissible scale transformation of the interval (\ref{14}). The obtained expressions for  $\Sigma(\phi)$ shows that the non-minimal coupling in our model reduces to a minimal one, if the target space is a sphere, a plane or a pseudo-plane (Lobachevsky space).  The model with such a degeneration of non-minimal coupling mains unchanged compared with a standard nonlinear sigma model, and can be involved in the investigation of dark energy and dark matter, as it made in Ref. \cite{C31}. Therefore, in what follows we assume that the condition of degeneration of non-minimal coupling is not fulfilled.

The non-zero connection coefficients (\ref{8}) for (\ref{15}) take the form:
\begin{equation}\label{17}
{\cal T}^{1}_{11}=\frac{Q'}{2Q},\,\,{\cal T}^{1}_{22}=-\frac{\epsilon \Sigma}{2Q}(\Sigma Q' + 2 Q \Sigma '),\,\,{\cal T}^{2}_{12}={\cal T}^{2}_{21}=\frac{1}{2 Q \Sigma} (\Sigma Q' + 2 Q \Sigma ').
\end{equation}
In view of expressions (\ref{15}) and (\ref{17}), the basic equations of the model (\ref{11}) - (\ref{13}) can be written as follows:
\begin{equation}\label{18}
3M_P ^{2}H^2=  \frac{1}{2} Q \Big(\epsilon \dot\phi^2 + \Sigma^2 \dot\chi^2\Big) + V(\phi,\chi),
\end{equation}
\begin{equation}\label{19}
M_P ^{2}\dot H=-  \frac{1}{2} Q \Big( \epsilon\dot\phi^2 + \Sigma^2 \dot\chi^2\Big) ,
\end{equation}
\begin{equation}\label{20}
\ddot\phi+3H\dot\phi+\frac{Q'}{2Q}\dot \phi^2-\frac{\epsilon \Sigma}{2Q}(\Sigma Q' + 2 Q \Sigma ')\,\dot\chi^2
+\frac{\epsilon}{Q}\frac{\partial V}{\partial\phi} = 0,
\end{equation}
\begin{equation}\label{21}
\ddot\chi+3H\dot\chi+\frac{1}{ Q \Sigma} (\Sigma Q' + 2 Q \Sigma ')\,\dot\chi\dot\phi
+\frac{1}{Q\Sigma^2}\frac{\partial V}{\partial\chi} = 0.
\end{equation}
From these equations, it is clear that the presence of the factor $ Q(\Sigma) $, which is  created by the non-minimal coupling and capable, in general, to change its sign, leads to an interesting phenomenon.We have in mind the possibility due to this factor for the fields $\phi$ and $\chi$ to change its character from quintessence to phantom, and vice versa . Moreover, this switching of the field type  occurs at the same time for both of them.

\subsection{\large Case of a constant potential}

Equations (\ref{18}) - (\ref{21})  can be easily solved in the case of constant potential (cf. Ref. \cite{C31}). Indeed, by integrating Eqs. (\ref{20}), (\ref{21}) for $V = V_0 = const.$, one can obtain two first integrals in the form of kinetic terms $K_{\chi}$ and $K_{\phi}$ for $\chi$ and $\phi$, respectively:
\begin{equation}\label{22}
K_{\phi} = \epsilon \frac{1}{2}Q \dot \phi^2 = \epsilon\frac{ C_0}{a^6}-\frac{C}{a^6 Q \Sigma^2},\,\,\,K_{\chi} = \frac{1}{2}Q\Sigma^2 \dot \chi^2 = \frac{C}{a^6 Q \Sigma^2},
\end{equation}
where $C,\,C_0$ are the constants of integration. Substitution of the kinetic terms from  (\ref{22}) into Eqs.  (\ref{18}), (\ref{19}) leads them to the form
\begin{equation}\label{23}
3M_P ^{2}H^2=  \epsilon\frac{ C_0}{a^6} + V_0,\,\,\,M_P ^{2}\dot H = -  \epsilon\frac{ C_0}{a^6}.
\end{equation}
Note that the second equation in (\ ref {23}) is simply a consequence of the first differential equation. Integrating the first equation for the scale factor $a(t)$, we obtain, as in \cite{C16}, that
\begin{equation}\label{24}
a(t) = a_0 \left[\sinh \left( \sqrt{\frac{3V_0}{M^2_{P}}}\,t\right)\right]^{\displaystyle \frac{1}{3}} \Rightarrow H(t) = \sqrt{\frac{V_0}{3 M^2_{P}}} \coth\left( \sqrt{\frac{3V_0}{M^2_{P}}}\,t\right),
\end{equation}
where $\epsilon C_0 = a_0^6 V_0 > 0$. It is interesting to noted that there is an unlimited number of ways of realization such a model, since the function $\Sigma (\phi)$ in the equations (\ref{22}) for the chiral fields remains arbitrary. Besides, $Q(\Sigma)$ is determined via the same function. Therefore, the class of  allowable metrics (\ref{14}) should be restricted with the help of a certain additional requirement, such as following from the perturbation analysis or from a stability condition.

\subsection{\large  The case of $V=V(\phi)$}

Let us consider one more illustrative example, continuing to follow the ideas of \cite{C31}. We now assume that the potential depends on the field $\phi$ only, that is $V=V(\phi)$. In this case, equation (\ref{21}) possesses the same first integral, as in Eq. (\ref{22}):
\begin{equation}\label{25}
\dot \chi^2 = \frac{2 C}{a^6 Q^2 \Sigma^4}.
\end{equation}
As two independent equations among three remaining equations (\ref{18}) - (\ref{20}), Eqs. (\ref{18}) and (\ref{19}) can be chosen, which due to Eq. (\ref{25}) can be now reduced to the form:
\begin{equation}\label{26}
3M_P ^{2}H^2= \epsilon \frac{1}{2} Q  \dot\phi^2 + \frac{C}{a^6 Q \Sigma^2} + V(\phi),
\end{equation}
\begin{equation}\label{27}
M_P ^{2}\dot H= -  \epsilon \frac{1}{2} Q  \dot\phi^2 - \frac{C}{a^6 Q \Sigma^2} .
\end{equation}
Last  equation of the system (\ref{18}) - (\ref{21}), i.e.  Eq.  (\ref{20}), is a consequence of Eqs. (\ref{26}) and (\ref{27}). Nevertheless, there are still several possibilities, which could be expressed with the help of some ansatz.

\subsubsection{Solutions from some ansatz}

First, note that the requirement of ansatz, similar to that used in Ref. \cite{C31},
\begin{equation}\label{28}
{\cal H}_{22}=Q \Sigma^2=a^{\displaystyle n-6},
\end{equation}
with $n$ being a real number, imposes a rather strong restriction on the allowable form of function $Q(\Sigma)$.Indeed, the positivity of the scale factor and the definition of $Q(\Sigma)$, according to Eq. (\ref{15}), lead to the non-equality $\epsilon\,\delta\Sigma\,''\le \alpha\Sigma$,
which is valid not for any function at any given time, and not for any values $\alpha, \beta$ and $\gamma$.  However, if we assume that this condition is satisfied, that is $Q = a^{n-6}/\Sigma^2$, and suppose the ansatz, similar to that used in \cite{C31} for the field $\phi(t)$ in the form
\begin{equation}\label{29}
\dot \phi^2 = \frac{2 B}{Q}=2 B a^{\displaystyle 6-n}\Sigma^2,
\end{equation}
where $B$ is some positive constant, one can write Eq. (\ref{27}) as follows:
\begin{equation}\label{30}
\dot H= -   \epsilon B   - \frac{C}{a^{\displaystyle n}}.
\end{equation}
Here and below, we set $M_P ^{-2} = 8 \pi G =1$, for the sake of simplicity.

The first integral of Eq. (\ref{30}) can be easily found :
\begin{equation}\label{31}
3 H^2 = -6\epsilon B \ln a +\frac{6 C}{n}\cdot\frac{1}{a^{\displaystyle n}}+\epsilon B +V_0,
\end{equation}
where the last two terms represent a constant of integration. Substitution of Eq. (\ref{31}) into Eq. (\ref{30}) enables us to write the expression for the potential as a function of the scale factor:
\begin{equation}\label{32}
V(a)=-6\epsilon B \ln a + \frac{6-n}{n}C a^{\displaystyle -n}+V_0.
\end{equation}
Equation (\ref{31}) can be written in terms of the variable $x(t)=\ln a(t)$,
\begin{equation}\label{33}
3 \,\dot x^2 = \epsilon B (1-6 x)+\frac{6 C}{n}\, e^{\displaystyle -n x} + V_0,
\end{equation}
that can be integrated in an explicit form only for certain values of the constant parameters $n,\,B,\,C$ and $V_0$.
Thus, we arrive at Eq. (\ref{33}),  which determines the dynamics of the model and completely coincides with the corresponding equation of the cited work. Therefore, further analysis of this model is not particularly interesting. We have to note only that in this case the behavior of the fields $\phi$ and $\chi$ is significantly complicated by the equations (\ref{25}) and (\ref{29}).

For illustrative purposes, let us consider solutions which could be referred to two different epochs in the evolution of the universe. We first consider the model of early expansion, when $a(t)\to 0$ and $x \to -\infty$. Retaining only the second term on the right-hand side of equation (\ref{33}) and integrating it, we get an expected result:
\begin{equation}\label{34}
a(t)=\left[ \sqrt{\frac{n C}{2}}\,(t-t_p)+a_p^{\displaystyle \frac{n}{2}}\right]^{\displaystyle \frac{2}{n}}\,\Rightarrow \, H(t)=\frac{2 H_p}{n H_p\,(t-t_p)+2}.
\end{equation}
where $t_p$ stands for the present time, and the present values of the Hubble constant and scale factor are related as $H_p =\sqrt{2C/n}\,a_p^{-n/2}$.

It is easy to find that the model undergoes the accelerated expansion with a constant deceleration parameter $q=-1-\dot H/H^2=-1+n/2$ once $n<2$. According to the formula  $V(t)=3H^2(t)+\dot H(t)$, one can find the potential of the field $\phi$ as follows
\begin{equation}\label{35}
V(t)=\frac{2 H_p^2(6-n)}{\Big[n H_p\,(t-t_p)+2\Big]^2}.
\end{equation}
In order to reconstruct the potential as a function of the field, it is necessary to find $\phi(t)$ from Eq. (\ref{29}). As can be seen, it is possible if, in addition to Eq. (\ref{28}) one assumes that for example $\Sigma^2=a^{-m}$. Substituting this expression into Eq. (\ref{29}) and using the solution (\ref{34}), we obtain after integration that
\begin{equation}\label{36}
\phi(t)=\frac{2}{6-m}\sqrt{\frac{nB}{C}}\, \left[\frac{1}{2H_p}\sqrt{\frac{2C}{n}}\Big( n H_p\,(t-t_p)+2 \Big)\right]^{\displaystyle \frac{6-m}{n}} + \phi_0,
\end{equation}
where $\phi_0$ is a constant of integration. Inserting Eq. (\ref{34}) into Eq. (\ref{35}), one can finally reconstruct the potential $ V(\phi)$, that supports the given  law of evolution of the scale factor:
$$
V(\phi) = \frac{\tilde{V}_0}{(\phi -\phi_0)^{\displaystyle 2n/(6-m)}},\,\,\mbox{where}\,\,\tilde{V}_0=C\frac{(6-n)}{n}\left[\frac{4nB}{(6-m)^2C}\right]^{n/(6-n)}.
$$
It can be seen that $V(\phi)\propto (\phi -\phi_0)^{-2}$, provided that $n+m =6$. But then from  Eq. (\ref{28}), it follows that $Q(\Sigma)=1$, i.e.  $\Sigma$ satisfies Eq. (\ref{16})  and takes the explicit form which is given above. Otherwise, when $n + m \neq 6$, one can find from equation (\ref{28}) and $\Sigma ^2 = a^{-m}$  that $\Sigma$ satisfies the following equation
\begin{equation}\label{37}
\epsilon\, \delta \,\Sigma\, \Sigma '' -\alpha \Sigma^2+\Sigma^{2(6-n)/m}=0.
\end{equation}
The possibility of finding exact solutions for this equation depends essentially on the exponent $2(6-n)/m$. If $n+m =6$, then we have $2(6-n)/m = 2$. This case has already been discussed above. Let $2n + m = 12$, that is, for example, $n = 3$ and $m = 6$. In this case, the solution of Eq. (\ref{37}) gives
$$
\Sigma(\phi) = \left\{
\begin{array}
[c]{rcl}
C_1\sinh\left(\displaystyle \sqrt{\frac{\alpha}{\delta}}\phi+C_2\right)+\alpha^{-1} ~~~\mbox{for}~~\epsilon = +1\,,\\
C_1\sin\left(\displaystyle \sqrt{\frac{\alpha}{\delta}}\phi+C_2\right)+\alpha^{-1}~~~\mbox{for}~~\epsilon = -1\,,
\end{array}
\right.
$$
where $C_1,\,C_2$ are constants of integration. The difference between this result  and the similar result, obtained above, is obvious, since now $\Sigma ''/ \Sigma \neq const.$ due to the non-zero parameter $\alpha^{-1}$.

As a second example, let us consider our model in the contemporary age, that is $x \to 0$. If we write the series for $\exp(-n x)$ in powers of $x$ in the right-hand side of Eq. (\ref{33}),  and retain the terms up to the second order, we obtain as a result
\begin{equation}\label{38}
3 \,\dot x^2 = \Big(\epsilon B + \frac{6 C}{n}  + V_0 \Big)-6 (\epsilon B + C) x+ 3 n C x^2.
\end{equation}
Recall that here $B,\,C,\,V_0$ and $n$ are free parameters of the equation. So it is easy to show that the substitution
$$
V_0 = \frac{3(\epsilon B + C)^2}{n C}-\epsilon B - \frac{6 C}{n}
$$
leads Eq. (\ref{33}) to the following equation
\begin{equation}\label{39}
\dot x = \sqrt{n C}\,\,|\,x-D|,\,\,\mbox{where}\,\,D=\frac{\epsilon B + C}{n C}.
\end{equation}
The result of integration of this equation depends on the sign of $D$.
One can see that for $\epsilon = +1$ we always have $D>0$, but if $\epsilon = -1$, then the sign of $D$ depends  on the sign of $B-C$. Given the definition of $x(t) = \ln a(t)$ and with an appropriate choice of the constants of integration, we obtain the following solutions for Eq. (\ref{39}):
\begin{equation}\label{40}
a=\exp\left[D\Big(1-e^{\displaystyle \mp \sqrt{nC}\,(t-t_p)}\Big)\right]\, \Rightarrow \, H= \pm \sqrt{nC}\,D \, e^{\displaystyle \mp\sqrt{nC}\,(t-t_p)},
\end{equation}
for $D>0$ and $D<0$, respectively. Keeping in mind that Eq. (\ref{39}) is written for $ x \to 0 $, i.e. $t \approx t_p$, we obtain from Eq. (\ref{40}) the following approximate solution: $a\approx \exp\{\sqrt{nC}\,|D|(t-t_p)\}$. According to this and the assumption that $\Sigma ^2 = a^{-m}$ in (\ref{28}), we obtain for the present epoch
$$
\phi(t) \approx \pm \frac{2\sqrt{2B}}{[6-(n+m)]D\sqrt{nC}}\, e^{\displaystyle \pm \frac{[6-(n+m)]}{2}D\sqrt{nC}\,(t-t_p)},
$$
where the constant of integration is equal to zero.
With the help of the latter, and $H(t)$ from Eq. (\ref{40}),  the field potential $V(\phi)$ can be approximately reconstructed  as
$$
V(\phi)\approx C D M\left( 3 D M \phi^{\displaystyle -2/[6-(n+m)]D}-1\right)\phi^{\displaystyle -2/[6-(n+m)]D},
$$
where
$$
M=\left[\frac{(6-n-m)^2nCD^2}{8B}\right]^{\displaystyle  -1/[6-(n+m)]D}.
$$

Let now the metric coefficient of the target space $h_{22} = a^{n-6} $, that is
\begin{equation}\label{41}
\Sigma(t)=a(t)^{\displaystyle \frac{n-6}{2}}.
\end{equation}
In Ref. \cite{C31}, $ n = 3 $ is chosen from the requirement for the field $\chi$  to mimic the pressureless dark matter with  equation of state $w_{dm} = 0$. Assuming that $\dot \phi^2 = 2 B$ (see also Ref. \cite{C31}) and making  use of $\Sigma'' = (1/\dot \phi^2) \ddot \Sigma - (\ddot \phi / \dot \phi^3) \dot \Sigma$, and Eqs. (\ref{15}), (\ref{41}),  we obtain the following expression:
\begin{equation}\label{42}
Q=\alpha - \epsilon \frac{\delta (n-6)}{8 B}[(n-6) H^2 + 2 \dot H].
\end{equation}
Under this, the dynamical equations (\ref{26}) and (\ref{27}) take the following form
\begin{equation}\label{43}
3 H^2= \epsilon B Q + \frac{C}{a^{\displaystyle n} Q} + V(\phi),
\end{equation}
\begin{equation}\label{44}
\dot H= - \epsilon B Q - \frac{C}{a^{\displaystyle n} Q} .
\end{equation}
Note that due to the the last  equations,  the expression (\ref{42}) could be written as
$$
Q=\alpha - \epsilon \frac{\delta (n-6)}{8 B}[(n-12) H^2 +2 V(\phi)].
$$

Thus, equation (\ref{44}) with $Q$ from Eq. (\ref{42}) is written as a nonlinear ordinary differential equation for $a(t)$ of the second order . Unfortunately, it is impossible to solve analytically this equation thus and, therefore, complete the reconstruction. Of course, there are other methods for reconstructing the field potential (see, for example, \cite{A17}, \cite{A18} and the references therein).

The substitutions, which we used above, were used in the cited papers, and were motivated by the mentioned reasons . It is hard to consider them quite successful in terms of perspective for obtaining exact solutions. Even  the solutions obtained here should be regarded as illustrative, since they have not undergone any tests, such as stability.
It could be noted that solving of Eq.  (\ref{44}) together with Eq. (\ref{42}) is a highly difficult problem due to its non-linearity.  One can specify exact solutions for this equation only for some particular values of its parameters.

\subsubsection{Approximation of a week coupling}

Let us study our model proceeding from a condition of the weak non-minimal coupling
This condition can be expressed as a small value of the second term as compared with the first term in Eq. (\ref{15}) for $Q(\Sigma)$,
\begin{equation}\label{45}
\left |\,\delta\frac{\Sigma''}{\Sigma}\,\right | \ll |\alpha|.
\end{equation}
Because of this inequality, the total kinetic term in Eqs. (\ref{26}), (\ref{27}) can be represented as follows
\begin{equation}\label{46}
K=K_{\phi}+K_{\chi}\approx \epsilon \frac{\alpha}{2} \dot\phi^2 + \frac{C}{\alpha \,a^6 \Sigma^2}+\epsilon\delta\Big(-\epsilon \frac{1}{2}\dot\phi^2 + \frac{C}{\alpha^2\,a^6 \Sigma^2}\Big)\frac{\Sigma ''}{\Sigma},
\end{equation}
and Eq. (\ref{27}) acquires the form
\begin{equation}\label{47}
\dot H = - \epsilon \frac{\dot\phi^2}{2}\left(\alpha -\epsilon\delta \frac{\Sigma ''}{\Sigma}\right)  - \frac{C}{\alpha^2 \,a^6 \Sigma^2}\left(\alpha +\epsilon\delta \frac{\Sigma ''}{\Sigma}\right).
\end{equation}
Being motivated by the reasons mentioned above, we require that the second term here is proportional to $1 / a^n$.  Absorbing the proportionality factor by $C$, this means  according to Eq. (\ref{27}) that
\begin{equation}\label{48}
\alpha +\epsilon\delta \frac{\Sigma ''}{\Sigma}= \alpha^2 \,a^{6-n} \Sigma^2.
\end{equation}
Taking into account the latter, equation (\ref{47}) can be rewritten as
\begin{equation}\label{49}
\dot H = - \epsilon \frac{\dot\phi^2}{2}\left(\alpha -\epsilon\delta \frac{\Sigma ''}{\Sigma}\right)-\frac{C}{a^n},
\end{equation}
which should be solved together with the equation (\ref{48}). If we substitute the scale factor from Eq. (\ref{48}) into Eq. (\ref{49}), and then use the inequality (\ref{45}), we can obtain the following equation
\begin{equation}\label{50}
\dot H = - \epsilon \frac{\dot\phi^2}{2}\left(1 -\epsilon\delta \frac{\Sigma ''}{\Sigma}\right)-C \Sigma ^{\displaystyle \frac{2n}{6-n}}\left(1 -\frac{\epsilon \delta\, n}{6-n} \frac{\Sigma ''}{\Sigma}\right),
\end{equation}
where we take $\alpha = 1$ without losing the generality. It is interesting that just in the case of $n=3$, expressions in the parentheses are the same, and this equation can be written as
\begin{equation}\label{51}
\dot H = - \left(\epsilon \frac{\dot\phi^2}{2} + C \Sigma ^2\right)\left(1 -\epsilon\delta \frac{\Sigma ''}{\Sigma}\right).
\end{equation}
Note that this equation must be solved together with Eq. (\ref{48}). Consequently, we have two equations for three unknowns: $a(t),\,\Sigma(\phi)$ and $\phi(t)$. Therefore,  it is necessary to have an additional condition on one of these functions or establish some relations between them. Here,  a large number of possibilities exists which may be realized in the framework of the  generating function method \cite{A19}, \cite{A21}, the method developed in  \cite{A22}, \cite{A23} etc.

\subsubsection{A new ansatz}

The use of substitutions of the type (\ref{28}) and (\ref{34}) is dictated by the expected property (the equation of state) of dark matter and imposes rather strict restrictions on the model. At the same time, the absence of sufficient number of equations, as compared with the number of parameters, and because of their nonlinearity,  it is obvious that some  additional restriction on the set of equations should be presented. There is a number of possibilities in addition to those that have been used above and in Refs. \cite{C27}, \cite{C30}, \cite{C31}. Let us consider an example of substitution, which in some extent generalizes the ansatz (\ref{28}) for $\phi$, and makes it easy to restore the dependence $\Sigma(\phi)$. Let us assume that \begin{equation}\label{52}
Q(\Sigma) \equiv \alpha-\epsilon \delta\frac{\Sigma''}{\Sigma} = \left(\frac{d \Psi}{d \phi}\right)^2,
\end{equation}
where $\Psi(\Sigma)$ is a differentiable function $\Sigma(\phi)$. The motivation for this ansatz could be as follows. In view of Eqs. (\ref{14}), (\ref{15}) and (\ref{52}), the expression for the non-minimal metrics can be written  identically to the original one, but with the transformed first field, $\phi \to \Psi$, that is
$$
d s^2_{nonmin}\equiv Q(\Sigma)d^2_{\sigma}=\epsilon d \Psi^2+\left(\Sigma \Sigma ' \frac{d\Psi}{d \Sigma}\right)^2 d \chi^2,
$$
Substituting (\ref{52}) into Eq. (\ref{27}), we obtain the following equation
\begin{equation}\label{53}
\dot H= -  \epsilon \frac{1}{2} \dot \Psi^2 - \frac{C}{ \left(\displaystyle a^3 \frac{d\Psi}{d \Sigma} \,\Sigma\,\Sigma '\right)^2 }.
\end{equation}
This equation is playing the  crucial role in obtaining various cosmological solutions.
The arbitrariness of function $\Psi(\Sigma)$ provides a great opportunity to study the model. Moreover, it allows us to find the explicit form for the metric coefficient of the target space $\Sigma^2(\phi)$ as the solution of Eq. (\ref{52}). It is interesting that  the model studied above on the basis of the  equation (\ref{30}) may be repeated again. Indeed, assuming as in (\ref{28}) $Q \Sigma ^2 = a^{n-6}$, we arrive at the equation (\ref{30}). The only difference is that now the kinetic term is proportional to $\dot \Psi^2$, and therefore $\dot \Psi^2 = 2B$. However,  the ansatz (\ref{52}) is more attractive due to the fact that, by giving an explicit form of the function $\Psi(\Sigma)$, one may immediately obtain $\Sigma(\phi)$ as the solution of Eq. (\ref{52}). Of course, this equation can not always be solved analytically. Nevertheless, if  it is possible, a one more condition is required for the set of equations (\ref{52}), (\ref{53}). Here, there are also various approaches to the problem.

One can require that the first term on the right-hand side of equation (\ref{53}) is a constant, that is to set again $\dot \Psi^2 = 2B$. From the latter, the dependence $\Sigma(t)$ for a given function $\Psi(\Sigma)$  follows immediately. Then, from the integral of Eq.  (\ref{52}) one can find $\psi(t)$. After that, the time dependence of $(d\Psi/d \Sigma) \,\Sigma\,\Sigma '$ could be derived. Thus, only one unknown function remains in Eq. (\ref{53}), e.g.   the scale factor $a(t)$.  We consider a simple example. Let
\begin{equation}\label{54}
\Psi(\Sigma)=\ln \Sigma.
\end{equation}
Then Eq. (\ref{52}) is written as
\begin{equation}\label{55}
\epsilon\, \delta\, \Sigma\, \Sigma\, '' +\Sigma\,'^2 - \Sigma^2=0,
\end{equation}
where we consider $\alpha=1$.

a) Let us restrict our study in this section to the case $\epsilon=+1,\,\delta \neq 0$. Then a particular solution of this equation can be written as
\begin{equation}\label{56}
\Sigma(\phi)=\left[\sinh\left(\frac{\sqrt{\phantom{\dot A}\!\!\!\delta+1}}{\delta}\,\phi\right)\right]^{\displaystyle \frac{\delta}{\delta+1}}.
\end{equation}
Next, we need an additional condition. For example,  one of the conditions discussed above can now be employed as well. What is interesting, we can set for the free parameter $\delta$ such a value that leads to a simplest form of argument in (\ref{56}). It means that we let  the factor $\sqrt{\delta+1}/\delta =1$. One can verify that the only positive value of parameter obeying this equation equals to $\delta=\delta_{gr}=(1+\sqrt{5})/2$, that is the so-called "golden ratio". However, we will proceed by a different way desiring to simplify our equations.

In order to simplify what follows, we can set for the free parameter $\delta$ such a value that leads to a simplest form of Eq. (\ref{55}): $\delta = 1$. In view of that, instead of Eq. (\ref{56}) we have  $\Sigma(\phi)=\sqrt{\sinh (\sqrt{2}\phi)}$.

If we now require that $\dot \Psi^2 = 2B\,\,\Rightarrow\,\,\Psi(t)=\sqrt{2B}\,t$, then from (\ref{54}), we obtain the following dependence of the original field on time
$$
\phi(t)=\frac{1}{\sqrt{2}}\sinh^{-1} \left[\exp\left(\displaystyle 2 \sqrt{2B}\,t\right)\right].
$$
From the last equations, it can be found that
$$
\left(\frac{d\Psi}{d \Sigma} \,\Sigma\,\Sigma '\right)^2 =\Sigma '\,^2=\cosh\left(2 \sqrt{2B}\,t\right).
$$
Substitution of this expression and $\dot \Psi^2 = 2B$ into Eq. (\ref{53}) leads to the following equation for the scale factor
\begin{equation}\label{57}
v \ddot v-\dot v^2 + 3 B v^2=-3 \sec{\mbox{h}}\left(2 \sqrt{2B}\,t\right),
\end{equation}
where $v=a^3$. Obtaining analytic solution of this equation is rather problematic. However, we have to note that the substitution (\ref{54}) hardly implies the existence of exact solutions from the very beginning. In any case, equation (\ref{57}) can be solved graphically by setting the initial data for $v$ and $\dot v$ and certain numerical values of the parameters in this equation. Finally, note that the potential is determined by the following simple formula, $V=\ddot v / 3 v$, and the Hubble parameter is defined as $H=\dot v/3v$.

b) Now we consider the case $\epsilon=+1,\,\delta = -1$, that is equivalent to $\epsilon=-1,\,\delta = 1$ with respect to Eq. (\ref{55}). Under this condition, the peculiar solution of Eq. (\ref{55}), with zero-valued constants of integration, is given by
\begin{equation}\label{58}
\Sigma(\phi)=\exp\Big(-\frac{1}{2}\phi^2\Big).
\end{equation}
Here,we suppose that instead of constancy of the first term in the right-hand side of equation (\ref{53}) we assume again that the second term is proportional to $1 / a^n$. Then, according to (\ref{54}), we have
\begin{equation}\label{59}
\left(\frac{d\Psi}{d \Sigma} \,\Sigma\,\Sigma '\right)^2 =\Sigma '\,^2=a^{\displaystyle n-6}.
\end{equation}
Taking into account this equation and Eq. (\ref{58}), we can find that
\begin{equation}\label{60}
\Sigma '\,^2 = \phi^2 e^{\displaystyle -\phi^2} = a^{\displaystyle n-6}.
\end{equation}
Therefore, the scale factor can be expressed in terms of the scalar field $\phi$ as
\begin{equation}\label{61}
a=\left(x e^{\displaystyle -x}\right)^{\displaystyle 1/(n-6)}\,\,\Rightarrow\,\,H=\frac{1}{(n-6)}\,\left(\frac{1}{x}-1\right)\,\dot x,
\end{equation}
where and below $x=\phi^2$. With the help of Eqs. (\ref{54}) and (\ref{61}), we can write the first term in (\ref{53}) as $(-\epsilon \dot \Psi^2)/2=-\epsilon \dot x^2/8$. With this and Eqs. (\ref{60}), (\ref{61}), we can finally represent Eq. (\ref{53}) as follows
\begin{equation}\label{62}
x(1-x)\ddot x +\left(\frac{\epsilon (n-6)}{8}x^2-1\right)\dot x^2 + C (n-6)x^2 \left(x e^{\displaystyle -x}\right)^{\displaystyle n/(6-n)}=0.
\end{equation}
It seems impossible to give an analytical solution to this equation at arbitrary values of $n$ and $C$. By applying the appropriate method for solving this equation,  may be in some approximation, we could readily restore the scale factor and the Hubble parameter according to Eq. (\ref{61}). Let us turn to the study of this problem in our subsequent study. In the present paper, it is important for us to demonstrate how this approach works.

\section{\large  Discussion  and  Conclusion}

The choice of the target-space metric in the form Eq. (\ref{14})  represents only a peculiar but of course allowable form of the generic two-dimensional metric.
As is known \cite{A24}, any two-dimensional (pseudo-)Riemannian space has to be conformal with the (pseudo-)Euclidean space. Therefore, we can write the target metric in the following form
\begin{equation}\label{63}
d s_{\sigma}^2 = \exp[2 S(\phi,\chi)](\epsilon d \phi ^2 +  d \chi^2).
\end{equation}
From this, we can readily derive the following non-zero components ${\cal R}_{AB}$ and ${\cal R}$:
\begin{equation}\label{64}
{\cal R}_{AB}=diag\left(-\epsilon\triangle_{\displaystyle \,\epsilon} S , - \triangle_{\displaystyle \,\epsilon} S \right),\,\,\,\,{\cal R}=-2\,e^{\displaystyle 2 S}\,\triangle_{\displaystyle \,\epsilon} S ,
\end{equation}
where
$$\triangle_{\displaystyle \,\epsilon} S=\epsilon\, \partial^2 S/\partial \phi^2 + \partial^2 S/\partial \chi^2.
$$
Therefore, we again arrive at ${\cal H}_{AB} = Q(S)\,h_{AB}$, where
\begin{equation}\label{65}
Q(S)=\alpha -\delta  \,e^{\displaystyle -2 S}\,\triangle_{\displaystyle \,\epsilon} S.
\end{equation}

Again, the non-minimal metric ${\cal H}_{AB} = Q(S)\,h_{AB}$. So, with a special choice of the coupling coefficients, $\alpha = 1,\,\, \gamma=-2\beta$,  the non-minimal coupling degenerates, that is, ${\cal H}_{AB}\equiv h_{AB}$ due to the vanishing of the Einstein tensor for any two-dimensional space. From Eq. (\ref{65}), it follows that the degeneration occurs, if the target function $S(\phi,\chi)$ obeys $Q(S)=1$, that is
\begin{equation}\label{117}
\triangle_{\displaystyle \,\epsilon} S =\frac{\alpha}{\delta}\,\,e^{\displaystyle 2 S}\,.
\end{equation}
In the allowable case of  zero-valued parameter $\alpha$, this function $S(\phi,\chi)$ has to obey the Laplace equation, that is to be a harmonic function. An interesting problem could be considered even in the case on non-zero $\alpha$ but with such a harmonic function $S(\phi,\chi)$.
Obviously even more interesting perspective arises in the case of a non-minimal coupling with the target space of greater then two dimension, what is worth studying further.

In this paper, we investigated the role played in cosmology by the non-minimal coupling to the target space in non-linear sigma model. We have shown how this sort of coupling could modify the kinetic terms of the chiral fields and the Friedmann equations. For the sake of simplicity, we focused on the case of two-component sigma model.

In this preliminary study, we made two important points. We have found that the well known non-linear sigma model can get a natural modification due to its intrinsic geometry.In some illustrative examples, we have shown who our new model could work. Obviously, any other choice of ansatz can be investigated via the same way. For this end, we have applied several substitutions for the target metric both used in previous papers by other researchers and a new ansatz proposed here. Moreover, we have defined what is the acceptable degree of freedom in the choice of the non-canonical kinetic terms in the chiral space only through internal resources of the space.

\end{document}